\documentclass[fleqn,usenatbib]{mnras}

\usepackage{newtxtext,newtxmath}
\usepackage[T1]{fontenc}
\DeclareRobustCommand{\VAN}[3]{#2}
\let\VANthebibliography\thebibliography
\def\thebibliography{\DeclareRobustCommand{\VAN}[3]{##3}\VANthebibliography}

\usepackage{natbib}
\usepackage{psfig}
\usepackage{amsmath}
\usepackage{hyperref}
\def\be{\begin{eqnarray}}
\def\ee{\end{eqnarray}}
\usepackage{comment}
\usepackage{graphicx,subfigure}
\usepackage{lipsum}
\usepackage{xcolor}
\usepackage{soul}
\usepackage{appendix}
\usepackage{enumerate}

\usepackage{float}
\usepackage[flushleft]{threeparttable}

\title[MSP single pulses]{Single--pulse studies of three millisecond pulsars}

\author[Palliyaguru et al.]{N. T. Palliyaguru,$^{1}$\thanks{E-mail: nipuni.palliyaguru@ttu.edu} B. B. P. Perera,$^2$ M. A. McLaughlin,$^{3,4}$ S. Os{\l}owski,$^5$ and
\newauthor
G. L. Siebert$^6$\\
\\ $^1$ Department of Physics and Astronomy, Texas Tech University, Lubbock, TX 79409, USA
\\ $^2$ Arecibo Observatory, University of Central Florida, HC3 Box 53995, Arecibo, PR 00612, USA
\\ $^3$ Department of Physics and Astronomy, West Virginia University, Morgantown, WV 26501, USA
\\ $^4$ Center for Gravitational Waves and Cosmology, West Virginia University, Chestnut Ridge Research Building, Morgantown, WV 26505, USA
\\ $^5$ Manly Astrophysics, 15/41-42 East Esplanade, Manly, NSW 2095, Australia
\\ $^6$ Department of Physics, University of Wisconsin Madison, Madison, WI 53703, USA
}

\date{Accepted XXX. Received YYY; in original form ZZZ}

\pubyear{2021}

\begin{document}
\label{firstpage}
\pagerange{\pageref{firstpage}--\pageref{lastpage}}
\maketitle

\begin{abstract}
Single--pulse studies are important to understand the pulsar emission mechanism and the noise floor in precision timing. 
We study total intensity and polarimetry properties of  three bright millisecond pulsars -- PSRs J1022+1001, J1713+0747, and B1855+09 -- that  have detectable single pulses at multiple frequencies. 
We report for the first time the detection of single pulses from PSRs J1022+1001 and J1713+0747 at 4.5~GHz.
In addition, for those two pulsars the fraction of linear polarization in the average profile is significantly reduced at 4.5~GHz, compared to 1.38~GHz, which could support the expected deviation from a dipolar field closer to the pulsar surface. 
There is a hint of orthogonal modes in the single pulses of PSR~J1713+0747.
More sensitive multi--frequency observations may be useful to confirm these findings.
The jitter noise contributions at 1.38~GHz, scaled to one hour, for PSRs~J1022+1001, J1713+0747 and PSR~B1855+09  are $\approx$135~ns, $\approx$45~ns, and $\approx$60~ns respectively and are consistent with previous studies.
We also show that selective bright-pulse timing of PSR~J1022+1001 yields improved root-mean-square residuals of $\approx$22~$\mu$s, which is a factor of $\approx$3 better than timing using single pulses alone.
\end{abstract}


\section{Introduction}

In general, pulsar emission remains poorly understood. 
The commonly accepted model suggests that emission is produced by curvature radiation from bunches of charged particles moving along magnetic field lines at relativistic speeds.
The emitted photons split into electron--positron pairs, producing a cascade of secondary pair plasma \citep{gj69,rs75}.
Components of the electric field that are parallel ($\rm E_{\parallel}$) and perpendicular ($\rm E_{\perp}$) to the magnetic field lines produce highly linearly polarized emission.
Circular polarization is thought to be generated both by intrinsic mechanisms and/or by propagation effects within the magnetosphere \citep{m94}, where a time delay is introduced between $\rm E_{\parallel}$ and $\rm E_{\perp}$ due to  different refractive indices of the two orthogonal components in the magnetosphere \citep{gs90}.
As the pulsar beam crosses our line of sight, sweeping through different magnetic field lines, the polarization position angle (PPA), which is the angle between the magnetic field line and the fiducial plane (i.e., the plane passing through the rotation and magnetic axes) makes an S--shaped curve \citep{rc69}. The PPA information is utilized in constraining the geometry of pulsars using the rotating  vector model \citep[e.g.][]{ran83a,ran83b,mr11}. 

Millisecond pulsars (MSPs) are old neutron stars that are spun up to millisecond periods. They have smaller magnetospheres and relatively smaller ($\sim$four orders of magnitude) magnetic field strengths than canonical pulsars, which have not been through the recycling process \citep{lk12}.
This poses the question of whether emission of MSPs is different from canonical pulsars.
For example, the emission beam size of a canonical pulsar is correlated with its period but such a correlation cannot be clearly seen for MSPs \citep{kxl98}. In addition, MSP profiles appear to show less evolution with frequency than those of canonical pulsars \citep{kxl98}. Some properties of canonical and millisecond pulsars are similar, however. For instance, MSPs can emit giant pulses \citep{hc70}, like normal pulsars;
there is evidence that giant pulses are linked to high magnetic-field strengths at the pulsar light cylinder and to high-energy emission \citep{rj01,etk21}.
MSPs too may show a deviation from the dipolar field line structure close to the NS surface \citep[e.g.][]{kxl98,Gil2002,Kalapotharakos2021}.

Pulsar emission is extremely complex, requiring sophisticated models to explain observations that do not fit into the general picture. 
Average pulse profiles are stable in general, with some notable exceptions, such as PSRs J1713+0747 \citep{bkm18,Xu2021}, B1937+21 \citep{bkm18}, and PSR B1828-11 \citep{lhk10}.
However, emission is highly variable at a single--pulse level.
Phenomena such as giant pulses \citep{hc70}, mode changing \citep{bac70}, drifting sub-pulses \citep{dc68, bac73}, nulling \citep{ran86} and microstructure within single pulses \citep{cwh90} are not easily explained by a simple emission model.
The PPAs of many pulsars, in particular MSPs, deviate from the S--shaped sweep
and show discontinuities or unexpected jumps. These are sometimes separated by $90^\circ$, named orthogonal polarization modes (OPMs),
and may result from either two highly polarized orthogonal modes of emission \citep[e.g.][]{brc76,crb78} or propagation effects \citep{ms00}.
Even though the curvature radiation with bunches explains the radio emission of pulsars to some extent, it has serious drawbacks, including the inability to explain the production of bunches and their existence over the characteristic time of emission \citep{m92}. Furthermore, secondary pair plasma is insufficient to reach the charge density required by Maxwell's equations \citep[e.g.][]{m94}.
As alternative mechanisms, relativistic plasma emission {\bf \citep{mg99}}, anomalous Doppler emission {\bf \citep[e.g.][]{mu79}}, linear acceleration emission {\bf\citep{c73}}, and maser emission {\bf \citep{zs79}} have been proposed  \citep[see][for a review of these effects]{mel21}. However, none of these theoretical methods are capable of fully explaining the pulsar emission mechanism \citep{m94,mel21}. 
Single-pulse studies can provide valuable diagnostics of emission mechanisms. For example, Crab pulsar microstructures suggest coherent emission from strong plasma turbulence \citep{he07}. Also,  highly polarized single pulses have been used to distinguish between coherent curvature radiation and maser emission in the past \citep{mgm09}.
Therefore,
single--pulse studies, including single--pulse--polarimetry, of pulsars with a variety of properties are crucial to fully  understand pulsar emission physics \citep{ran86}. 

MSPs are also important for low-frequency, from nHz to $\mu$Hz, gravitational wave detection experiments through pulsar timing arrays (PTAs), which require precise measurements of pulse times of arrival (TOAs) and sub-microsecond timing accuracy \citep{aab+21,dcl+16,krh+20,pdd+19,psb+18}.
Pulse--to--pulse jitter, profile variations, variation in polarization properties, and polarization calibration errors are among phenomena that could affect the precision of TOAs \citep{cs10,Liu2011,osd13}.
These can contribute to noise that can limit the timing precision of MSPs and therefore sensitivity of PTAs to gravitational waves.
Therefore studying MSPs at a single--pulse level to understand the contributions of these various phenomena can offer useful insights.

Single--pulse studies of MSPs have been sparse due to signal--to--noise ratio limitations due to their low fluxes \citep{kxl98} and data acquisition requirements such as the  need for high time resolution sampling.
The few previous studies have revealed highly linearly polarized single pulses and sub--pulse microstructure in  J0437$-$4715 \citep{jak98,osb14,dgs16}, giant pulses from B1937$+$21 \citep{Mckee2019}, pulse jittering from PSR J1713$+$0747 \citep{sc12} and PSR J1022+1001 \citep{Liu2015, Feng2020}, and sub-pulse drifting from PSR J1713$+$0747 \citep{Liu2015}.

Polarization information in single pulses could get lost when averaging, causing depolarization in averaged pulse profiles \citep{br80}. Therefore studying single pulses may provide clues to the emission physics of pulsars.
In addition to emission mechanisms, there are other advantages to performing polarimetry.
\citet{b00} suggested the possibility of using invariant profiles to avoid errors due to calibration and \citet{s06} suggested the use of polarimetric profiles to improve timing.
Furthermore, \citet{osd13} used polarization information to correct for pulse--to--pulse variability in PSR J0437$-$4715, which resulted in a 40\% improvement in the timing precision of the pulsar.
In light of these ideas, we investigate single--pulse emission properties in three stable millisecond pulsars -- PSRs J1022+1001, J1713+0747, and B1855+09 -- which are monitored regularly by PTAs \citep[see][]{pdd+19}.
PSR~J1022+1001 is known to show long--term profile instabilities \citep[e.g.][]{kxc99,Liu2015,pbc21} and PSR J1713+0747 recently underwent a significant pulse shape change \citep[e.g.][]{Xu2021}. While the origin of such changes is likely intrinsic, and not due to incorrect calibration, these events can introduce variations in timing residuals, degrading the PTA sensitivity to gravitational waves.

The paper is structured as follows. The details of observations and data processing are presented in Section~\ref{sec:obs}. Data analysis techniques are presented in Section~\ref{sec:da} and results are presented in Section~\ref{sec:res}. The conclusion and discussion of the study are presented in Section~\ref{sec:sum}.

\section{Observations and data processing}
\label{sec:obs}

PSRs J1022$+$1001, J1713+0747, and B1855+09 were observed with the 305-m William E. Gordon Telescope at the Arecibo observatory in Puerto Rico. The basic parameters of these three millisecond pulsars\footnote{\url{https://www.atnf.csiro.au/research/pulsar/psrcat/}} are given in Table~\ref{tb:msplist}.
Arecibo observations were carried out between August 5, 2018 and August 25, 2019. 
Data were recorded using the Puerto Rico Ultimate Pulsar Processing Instrument (PUPPI) at center frequencies of 430 MHz, 1380 MHz, 2030 MHz and 4500 MHz
in full Stokes mode with 8--bit sampling.
Our multi-frequency observation details, including the usable bandwidth, number of channels across the usable bandwidth and the sampling time, for each pulsar, are listed in Table~\ref{tb:obs}.
The usable bandwidth at 1.38 GHz and 2.3 GHz is less than the full receiver bandwidth due to radio frequency interference (RFI).
While baseband recording mode allows higher time resolution, we accumulated data in search mode in order to maximize the available bandwidth and thereby to improve the S/N of the pulsar data. PSRs~J1713+0747 and B1855+09 data were coherently dedispersed using  \textsc{dspsr}\footnote{\url{http://dspsr.sourceforge.net}} \citep{vb11} at the dispersion measure listed in Table~\ref{tb:obs}.
PSR~J1022+1001 was observed in the incoherent search mode.
Each observation session started with a noise calibrator injection scan followed by the pulsar observation.

\begin{table}
\begin{center}
\caption{Properties of the observed pulsars: The pulsar period, period derivative, the dispersion measure, and references. 
}
\begin{tabular}{lcccc}
\hline
Name & Period & $\dot P$ & DM & Reference \\
&(ms)&($\rm s\,s^{-1}$)&$(\rm cm^{-3}\,pc$)\\
\hline
J1022+1001 &  16.453 & $4.334\times10^{-20}$&  10.252 & 1,2 \\
J1713+0747 & 4.570 & $8.530\times10^{-21}$ &15.917 & 3,4 \\ 
B1855+09 & 5.362 & $1.784\times10^{-20}$ & 13.314 & 4 \\
\hline
\label{tb:msplist}
\end{tabular}
\end{center}
\begin{tablenotes}
\small
\item References: (1)~\citet{hbo06}, (2)~\citet{Reardon16}, (3)~\citet{Zhu15}, (4)~\citet{Arzoumanina18}.
\end{tablenotes}

\end{table}

The recorded data were processed with \textsc{dspsr} \citep{vb11} to obtain single pulses, which were further processed with
\textsc{psrchive}\footnote{\url{http://psrchive.sourceforge.net}} routines \citep{hsm04}.
The data from each pulsar were processed with 
512 pulse phase bins given the time resolution used in the observation setup.
The automatic median zapping algorithms of \textsc{psrchive} were used to remove narrow--band and impulsive RFI.  Remaining RFI was removed by visual inspection.

Polarization calibration was performed to correct for the differential gain and phase between the two polarization channels of the receiver due to imperfections in the amplifiers and mismatch between the cable chains along the two paths, assuming an ideal feed.
For this, we used the calibrator scan which is a 25 Hz winking cal signal injecting a fully linearly polarized signal in between the two polarization probes at a 45$^{\circ}$ position angle. 
Next, the instrumental response was determined using polarization calibration modeling (pcm) as outlined in \citet{b00},
assuming equal ellipticities of receptors.
A flux calibrator was not observed during these observations. 
The polarization properties of the integrated profiles are shown in Figure~\ref{fig:singleprofs}, which are consistent with previously published profiles \citep{dhm15}.

Faraday rotation due to the magnetized ISM and the ionosphere  causes a change in the PPA with frequency \citep[e.g.][]{Simard80,Lyne89}. The change in the PPA due to Faraday rotation is given by 
\begin{equation}
    \Delta\Psi=\lambda^2\times \rm RM,
    \label{eq:rm}
\end{equation}
where $\lambda$ is the observation wavelength.
The rotation measure (RM), which depends on the average magnetic field and the electron density along the line--of--sight, was calculated using the \texttt{rmfit} program in \textsc{psrchive}, which searches for a peak in the linear polarization for trial RMs. 
The best-fit RM values (published values in parenthesis) are given in Table~\ref{tb:obs}.  Given the $\lambda^2$ dependence in Equation~\ref{eq:rm}, and the decrease in flux density with increasing frequency, the sensitivity to the change in the PPA decreases with increasing frequency. Therefore, the RM values derived from 1.38~GHz data were used to correct the data at higher frequencies.

Our RM values at 1.38~GHz are consistent within errors with published values for MSPs J1713+0747 and B1855+09 (see Table~\ref{tb:obs}). 
However, significant trends in the measured RM has been observed for PSR J1713+0747 previously \citep{Wahl21}. Our measured RM for PSR J1022+1001 differs significantly from its published value, however, 
PSR~J1022+1001 is known to show changes in the RM when passing close to the sun i.e. when the angular separation between the pulsar and the sun is $<3^\circ$ \citep{You11} and also shows significant long--term RM variations \citep {Yan11}.
The angular separation between the sun and the pulsar was $\approx15^\circ$ during the 1.38 GHz observations.
While it is not possible to pinpoint to the reason for the discrepancy between our measured and previously published RM values for PSR J1022+1001, we note that \citet{Noutsos2015}, \citet{Feng2020} and \citet{dhm15} also find higher RM values of 2.18(2), 2.9(2) and 4.68(6) $\rm rad\,m^{-2}$ respectively.

For the pulse phase jitter analysis, average profiles of 50, 100, 200, 500, and 1000 pulses were created. A TOA for each profile was generated by cross-correlating it against a noise-free template profile. This template was obtained by fitting Gaussian components to a high S/N profile obtained by averaging over the full observation for a given frequency. 
Timing residuals were obtained using the \textsc{tempo2} pulsar timing package \citep{Edwards06, Hobbs06}. The timing ephemerides are obtained from data published in \citet{pdd+19}.

\begin{table*}
\begin{center}

\begin{footnotesize}
\caption{Observation information, including MJD, frequency, observation length, bandwidth, number of channels across the bandwidth, sampling time, receiver temperature, sky temperature, measured rotation measures (published rotation measures are given in parentheses), 
equivalent width, measured mean flux density from the radiometer equation,  and the number (and fraction) of pulses detected  ($\rm S/N>5$). 
}
\resizebox{19cm}{!}{
\begin{tabular}{lllllllllllllll}

\hline
Name & MJD & $\nu$ & Length& BW & $\rm N_{chan}$& G &$t_{\rm samp}$&T$_{\rm rec}$ &T$_{\rm sky}$ & RM& W$_{\rm eq}$ & $S_{\rm mean}$ & N \\
&& (MHz)& (s) & (MHz) & &K/Jy & $(\mu $s) &(K) &(K)&$(\rm rad\, m^{-2})$ &(ms) &(mJy) &\\
\hline
J1022+1001 &  58335 & 1380 & 300& 600 & 384& 8 &20.48& 30 & 0.62 & $5.31\pm 0.04$ (--0.3, 2.18) &1.028$\pm$0.002& 4.89$\pm$0.15& 10755 ($\approx 59\%$) \\
& 58651&4500&1800 & 800 & 512 &4 & 40.96& 30&0.03 & --&0.577$\pm$0.003 &0.90$\pm$0.11 &1175 ($\approx 1\%$) \\
\hline
J1713+0747 & 58697 &1380 & 300& 600 &384& 8 &10.24& 30 &2.92 &$10.69\pm0.04$ (13$\pm$ 2) & 0.1752$\pm$0.0003 & 10.18$\pm$0.17  & 64458 ($\approx 98\%$) \\
& 58697& 4500 & 300 & 800 & 512 & 4 & 10.24 & 30 & 0.14&-- & 0.195$\pm$0.002& 1.37$\pm$0.27 & 53 ($\approx 0.08\%$) \\
\hline
B1855+09 & 58720& 430 & 600 &  20 & 64& 11 & 2.56& 50 &109.9&--&0.53$\pm$0.01& 10.61$\pm$2.15 & --\\
& 58720 & 1380 & 1200 & 600 & 384& 8 &10.24 & 30 &5.11&  $24.99\pm0.18$ (20$\pm$4) & 0.483$\pm$0.001&  5.21$\pm$0.09 &11480 ($\approx 10\%$) \\
&58720&2030 & 600 &460  &384& 8& 10.24&  40&1.94 &--& 0.43$\pm$0.02 & 0.39$\pm$0.17 & --\\
\hline
\label{tb:obs}
\end{tabular}
}
\begin{tablenotes}
      \small
      \item Notes:
      \item The published RM values are obtained from \citet{Yan11}, \citet{Noutsos2015} and \citet{Gentile2018}.
      \item The RM of PSR B1855+09 is not calculated at 430 MHz due to the small bandwidth.
      \item $ \rm T_{\rm rec}$ and Gain values for AO are obtained from  \url{http://www.naic.edu/~astro/RXstatus/rcvrtabz.shtml}.
      \item $ \rm T_{\rm sky}$ in the direction of the pulsar is calculated  according to \citet{Haslam82}. 
    \end{tablenotes}
\end{footnotesize}\end{center}\end{table*}
 
 
\section{Data analysis}
\label{sec:da}

In this section we discuss the data analysis tools used to measure total intensity properties, polarimetry of single pulses, and pulse phase jitter of PSRs J1022+1001, J1713+0747, and B1855+09.

Table~\ref{tb:obs} lists the measured equivalent pulse width $W_{eq}$ and the measured flux density $S_{mean}$ of the average profile over the full observation length from our multi-frequency observations. 
$W_{\rm eq}$ is defined as the width of a top hat pulse with the peak amplitude $I_{\rm peak}$ and the same area as the on--pulse region and is calculated as
\begin{equation}
    W_{\rm eq} = \frac{\sum_{i=n_1}^{N} I_i}{I_{\rm peak}},
    \label{eq:eqwidth}
\end{equation}
where $I_i$ is the intensity of the $i^{th}$ bin of the on--pulse region ranging from bins $n_1$ to N in baseline-corrected data. 
The error of $W_{\rm eq}$ is calculated by applying error propagation on Equation~\ref{eq:eqwidth} and using the off and on--pulse rms as the errors of $I_i$ and $I_{\rm peak}$ respectively.

From the radiometer equation, the root mean square noise fluctuation is given by 
\begin{equation}
    \Delta S_{\rm sys} = \frac{T_{\rm sys}}{\,G\sqrt{N_p\, t_{obs}\,\Delta \nu}}=\rm C \, \sigma_p.
    \label{eq:radiometer}
\end{equation}

\noindent
Here,  C is the scaling factor, $\sigma_p$ is the standard deviation of the off--pulse region, $T_{\rm sys}$ is the system temperature, which is the sum of the sky temperature ($T_{\rm sky}$), the receiver temperature ($T_{\rm rec}$), and the telescope spill over, $G$ is the telescope gain,  $\Delta \nu$ is the bandwidth, $t_{\rm obs}$ is the observation length, and $N_p=2$ is the number of polarization channels \citep[see][]{lk12}. 
The measured mean flux density $S_{\rm mean}$ is calculated by scaling the profile by $\rm C$ (i.e. $\Delta S_{\rm sys}/\sigma_p)$. The values for $T_{\rm rec}$, $T_{\rm sky}$, and $G$ for different frequencies are listed in Table~\ref{tb:obs}.

\subsection{Single--pulse properties}

\begin{figure*}
\includegraphics[width=18cm]{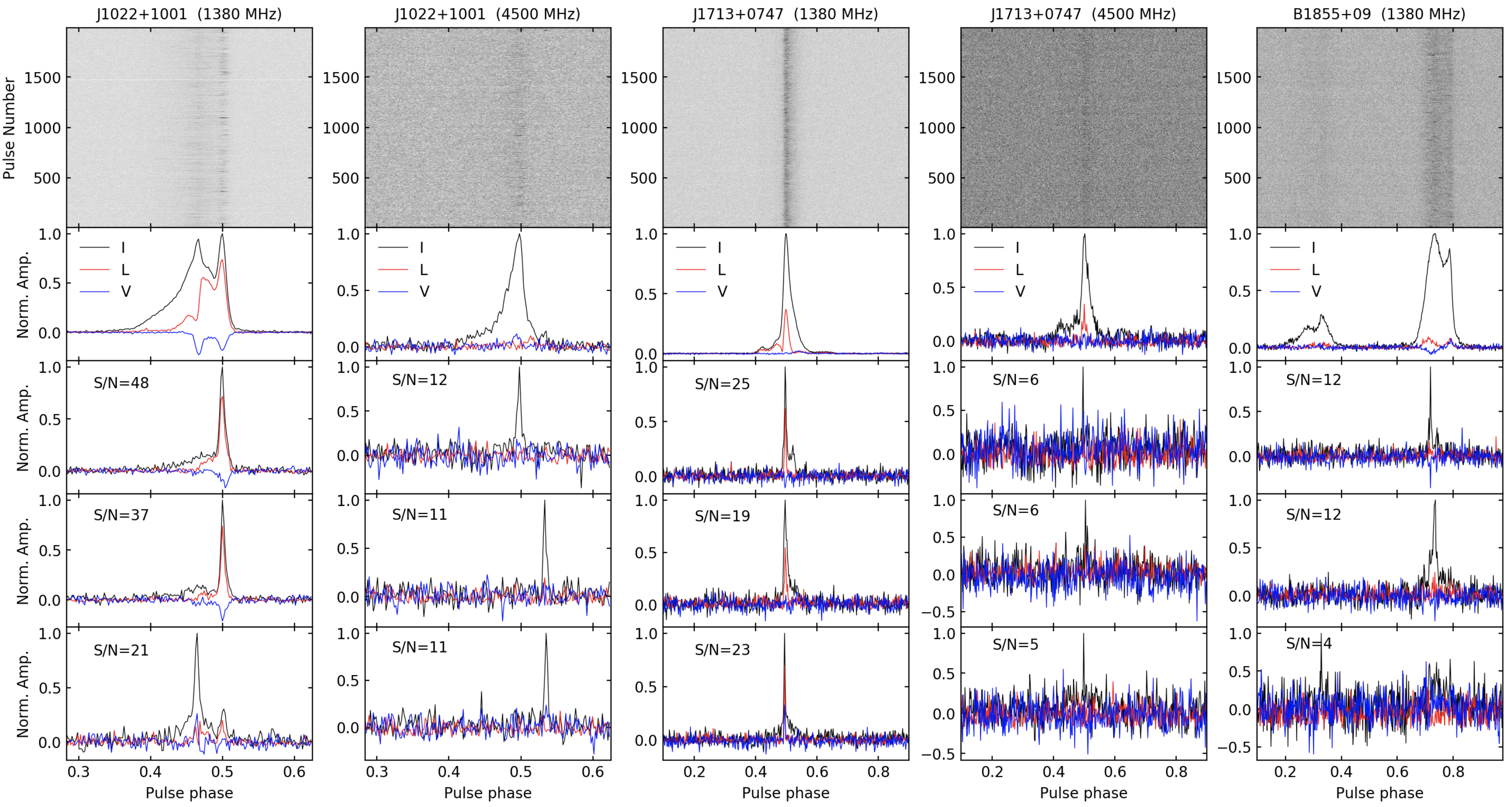}
\centering
\caption{
Multi-frequency single pulse observations of PSRs J1022$+$1001, J1713$+$0747, and B1855$+$09 at observing frequency 1.38 and 4.5 GHz (see different columns). For each pulsar at each frequency, a sequence of 2000 single pulses in grey scale ({\it top panel}) and the averaged profiles \text{of each pulse sequence}, including total intensity ({\it black}), linear ({\it red}) and circular ({\it blue}) polarization ({\it second panel from top}) are shown. Three highest S/N single-pulse profiles are shown from each observation ({\it bottom three panels}). A single pulse associated with the leading component of PSR~J1022+1001 at 1.38 GHz with a low fraction of linear polarization is also shown. Two single pulses of PSR~J1022+1001 at 4.5 GHz which appear at the small trailing component at phase $\approx 0.53$ (away from the main component) are also shown.
}
\label{fig:singleprofs}
\end{figure*}

We analyze single--pulse properties of  the five data sets.
Single-pulse amplitude and S/N distributions are used to identify phenomena such as giant pulses, which have flux density $>10\times$ the mean flux density \citep{ksl12,Knight07}.
The S/N for each single-pulse is calculated as

\begin{equation}
      S/N= \frac{I_{\rm peak}}{\sigma_p}
\end{equation}

\noindent where $I_{\rm peak}$  is the peak amplitude, 
and $\sigma_p$ is the off--pulse standard deviation of the pulse profile.
The number and the fraction of detected single pulses ($\rm S/N > 5$) is listed in Table~\ref{tb:obs}.

\begin{figure*}
    \begin{center}
    \hspace{-1.cm}
    \subfigure[]
{
\includegraphics[width=3.6cm,trim={ 0.35cm 0.0cm 0.4cm 0cm},clip]{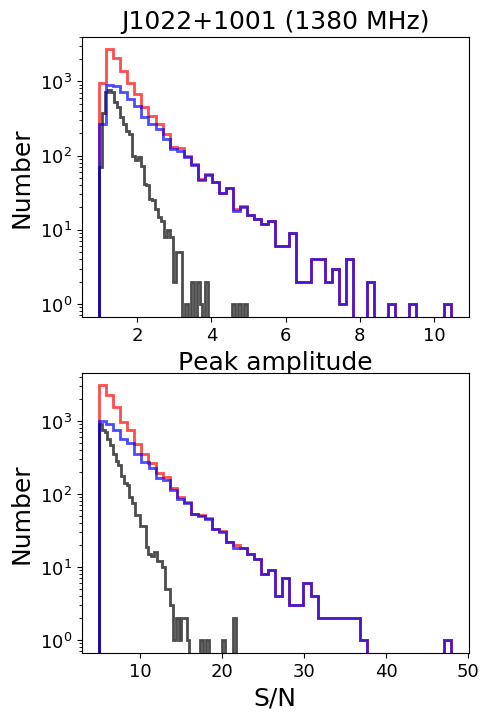} 
}
\subfigure[]
{
\includegraphics[width=3.45cm,trim={ 0.35cm 0.0cm 0.2cm 0cm},clip]{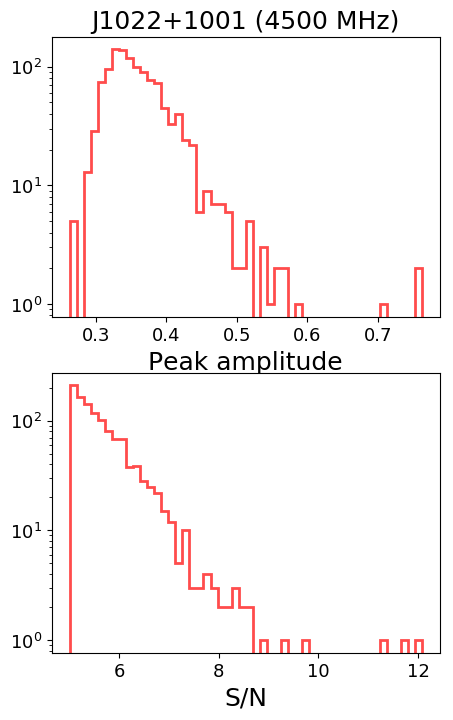} 
}
\subfigure[]
{
\includegraphics[width=3.45cm,trim={ 0.35cm 0.0cm 0.4cm 0cm},clip]{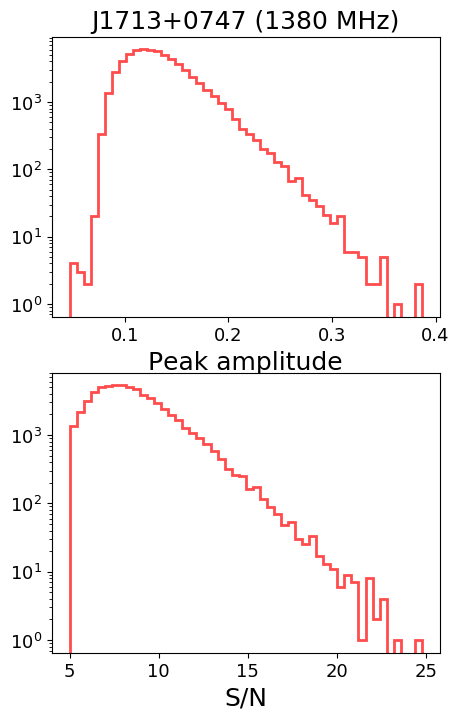} 
}
\subfigure[]
{
\includegraphics[width=3.6cm,trim={ 0.35cm 0.0cm 0.4cm 0cm},clip]{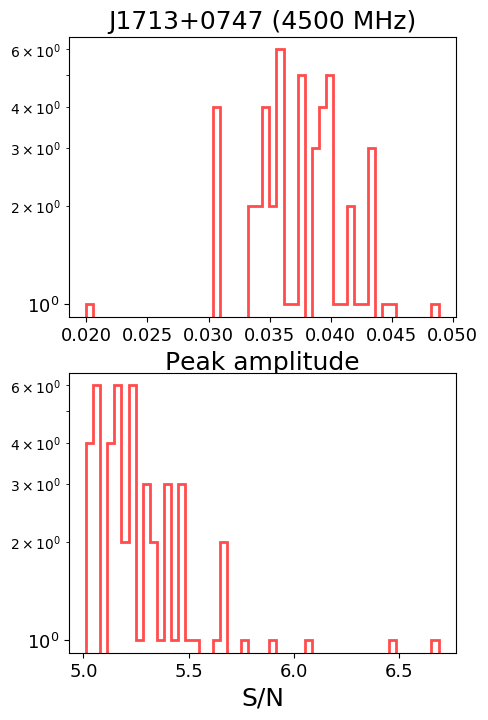} 
}
\subfigure[]
{
\includegraphics[width=3.4cm,trim={ 0.35cm 0.0cm 0.3cm 0cm},clip]{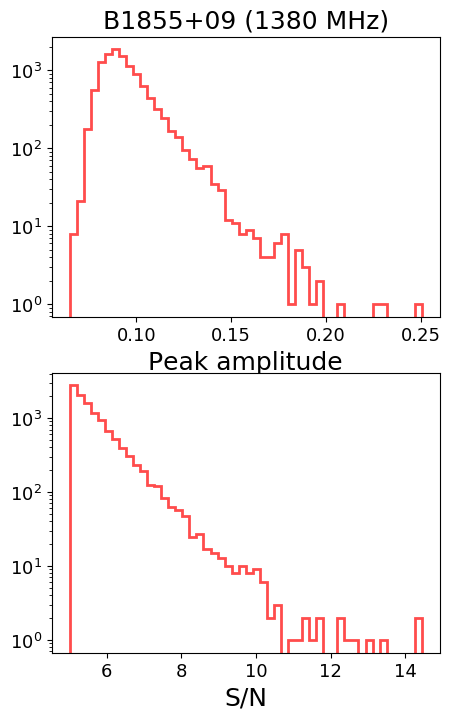} 
}
\caption{
Histograms of peak amplitude (top) and S/N (bottom) of PSR~J1022+1001 at 1.38 GHz (a) and 4.5 GHz (b), PSR~J1713+0747 at 1.38 GHz (c), and 4.5 GHz (d), and  PSR~B1855+09 at 1.38 GHz (e). For PSR~J1022+1001 at 1.38 GHz, the histograms of single--pulses associated with the leading (black) and trailing (blue) components are also shown separately in addition to the histograms of all single pulses (red). The brightest seem to correspond to the trailing component.
}
\label{fig:ampdist}
\end{center}
\end{figure*}

\begin{figure*}
    \begin{center}
    \hspace{-1cm}
    \subfigure[]
{
\includegraphics[width=4.7cm,trim={ 2.5cm 12.0cm 2.5cm 2.2cm},clip]{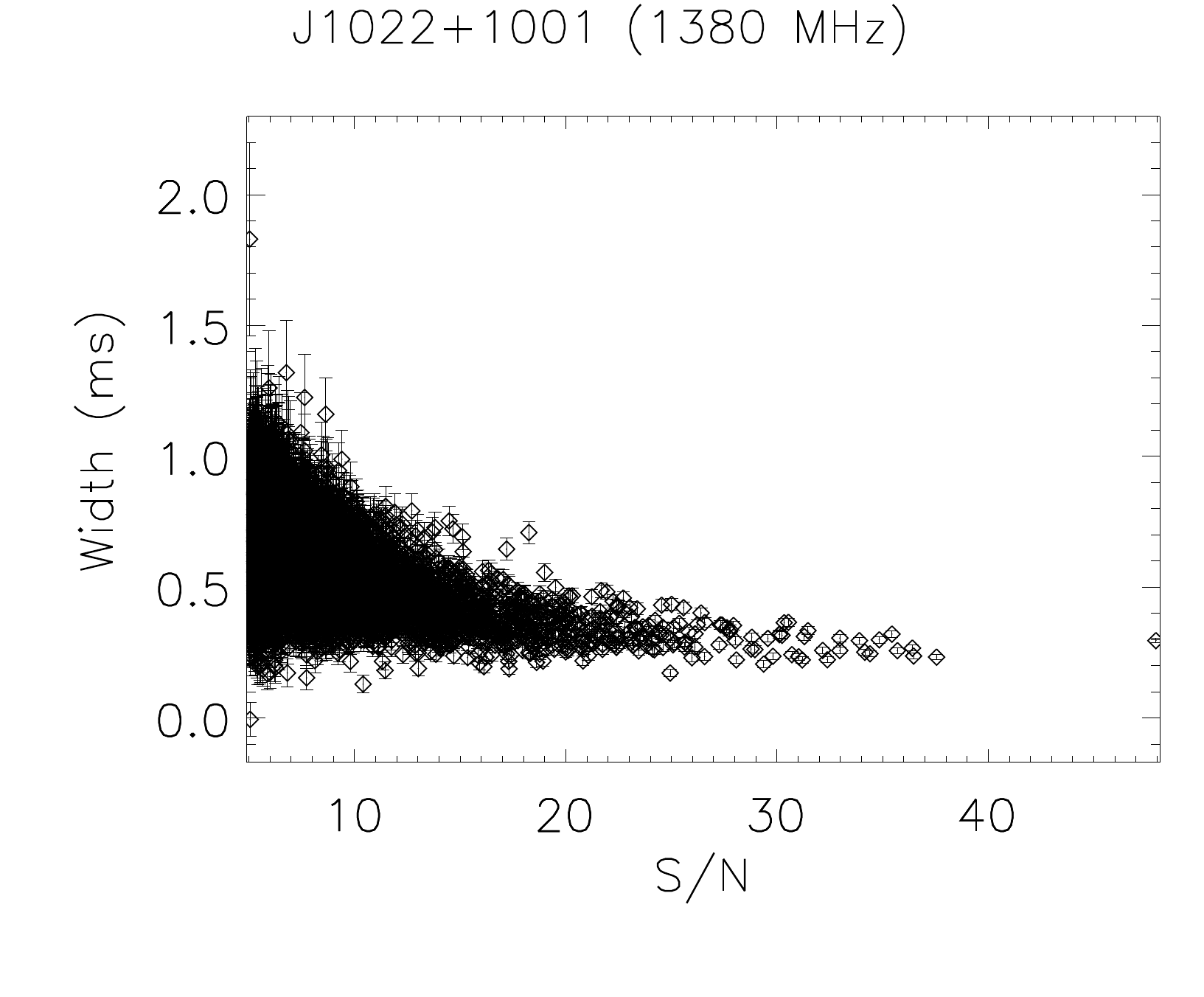} 
}
\subfigure[]
{
\includegraphics[width=4.3cm,trim={ 4.2cm 12.0cm 2.5cm 2.2cm},clip]{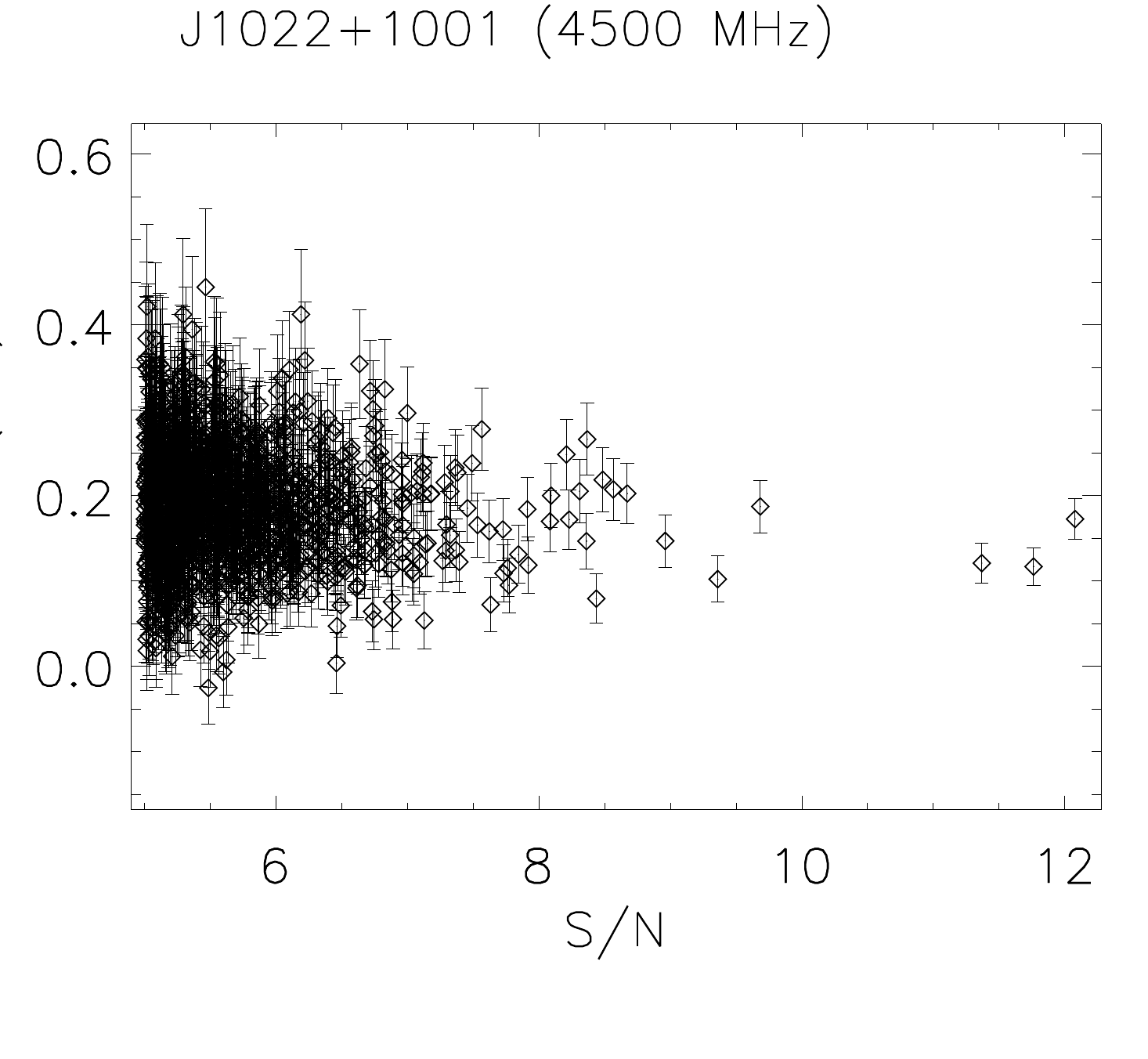} 
}
\subfigure[]
{
\includegraphics[width=4.3cm,trim={ 4.2cm 12.0cm 2.5cm 2.2cm},clip]{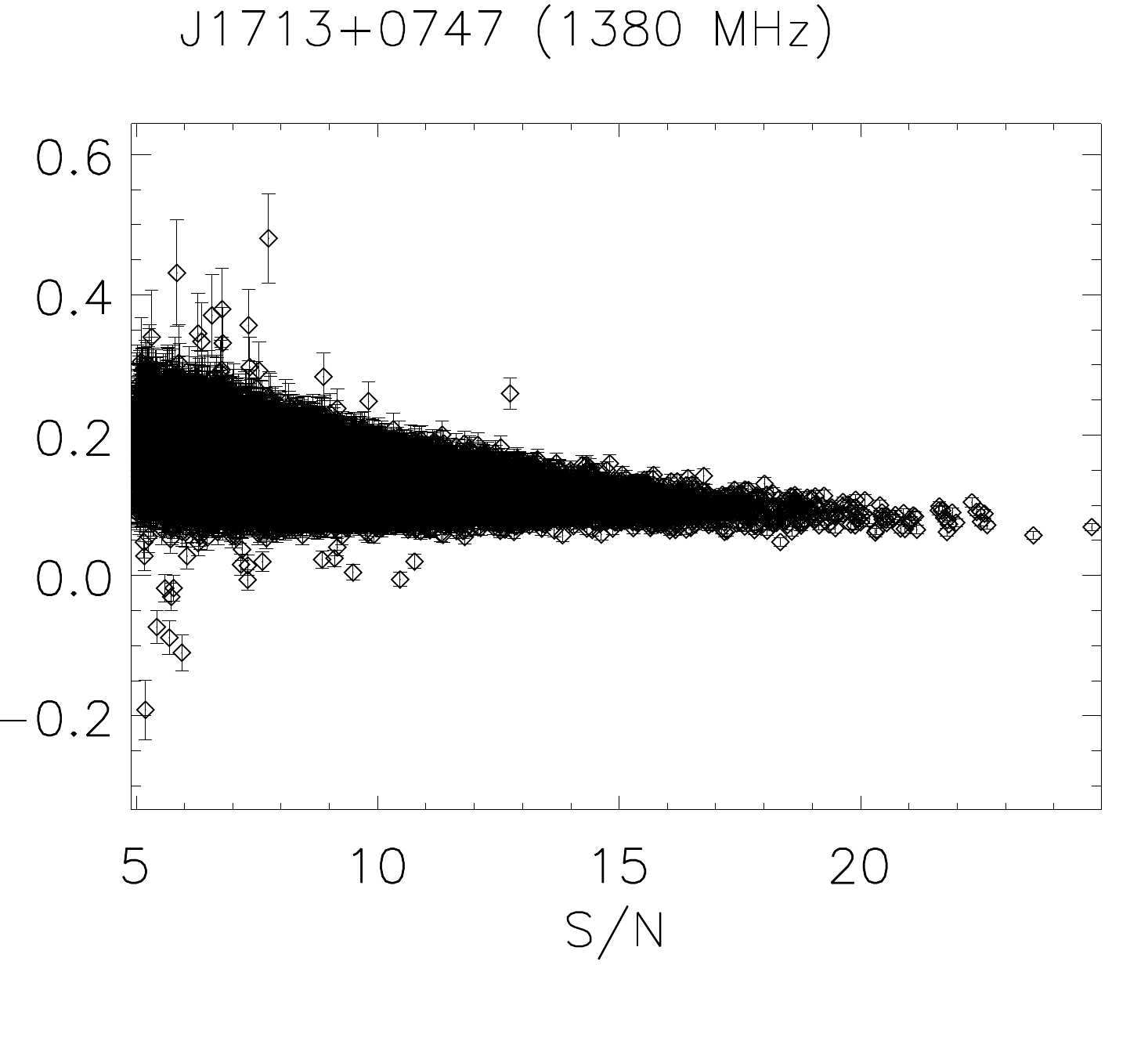} 
}
    \subfigure[]
    {
\includegraphics[width=4.3cm,trim={ 4.2cm 12.0cm 2.5cm 2.2cm},clip]{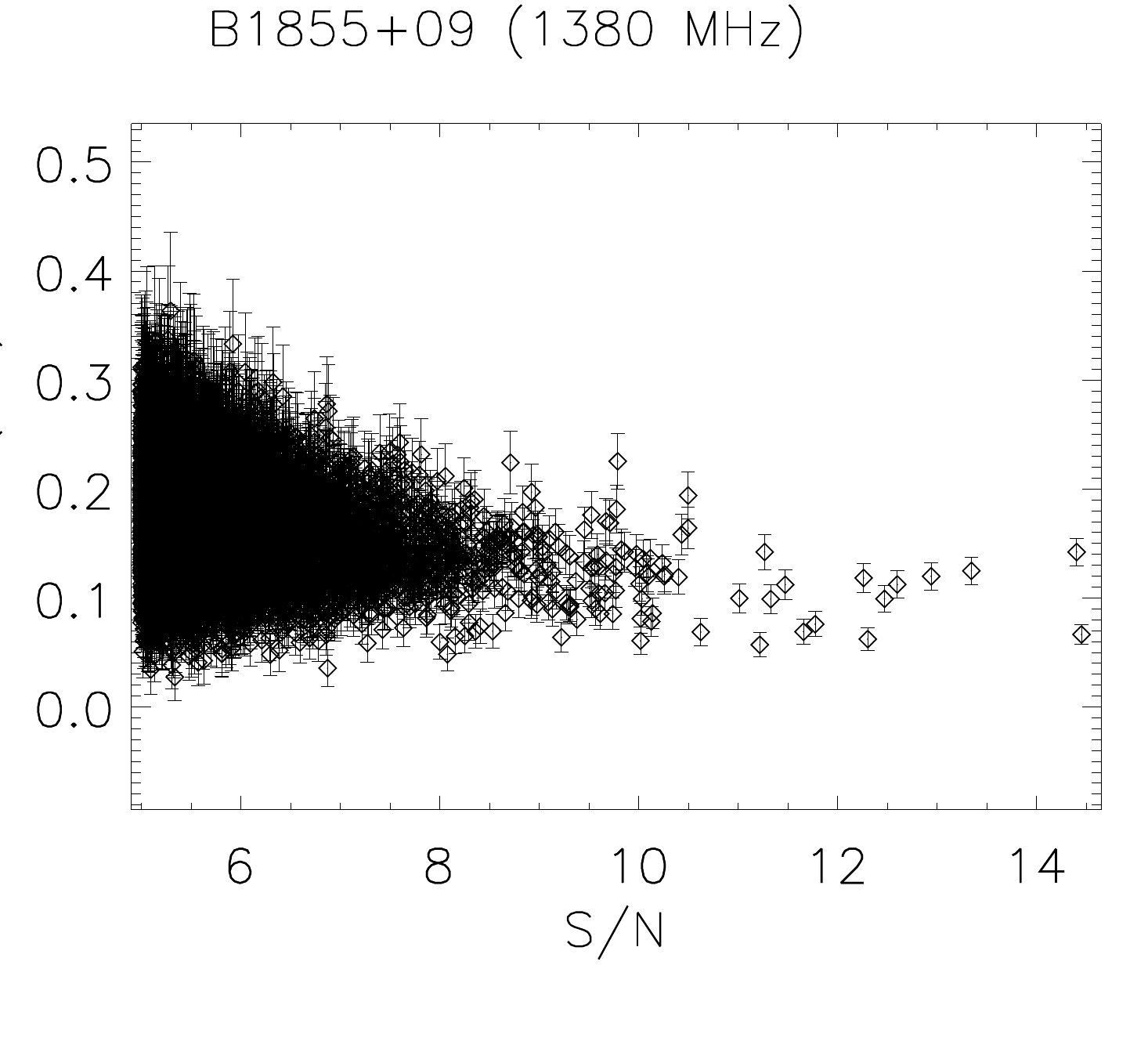} 
}
\caption{
Measured single--pulse equivalent width and S/N values for PSR~J1022+1001 at 1.38 GHz (a) and 4.5 GHz (b), PSR~J1713+0747 at 1.38 GHz (c), and  PSR~B1855+09 at 1.38 GHz (d). The S/N cutoff is used to show only measurements > $3\sigma$. Widths of PSR~J1713+0747 single pulses at 4.5 GHz are not shown due to the small fraction of detected pulses with S/N$>$5.
}
\label{fig:width_vs_snr}
\end{center}
\end{figure*}

\subsection{Single--pulse polarimetry}
\label{sp_pol}

Stokes parameters I, Q, U, V describe the polarization of electromagnetic waves, such that I is the total intensity, Q and U describe the intensity of linear polarization L as $L=\sqrt{Q^2+U^2}$, and V is the intensity of circular polarization \citep[see][]{lk12}.
The polarization position angle is defined as $\rm PPA = 0.5\tan^{-1}(U/Q)$.

\begin{table*}
\begin{center}
\begin{footnotesize}
\caption{Polarization parameters: percentage fractional linear and circular polarization of the average profile and the peak of the histograms of linear and circular polarizations shown in Figure~\ref{fig:linpol_single}. 
}
\begin{tabular}{llllll}
\hline
Pulsar & Frequency & $\rm L_{\rm av}/I_{\rm av}$ & $\rm V_{\rm av}/I_{\rm av}$ & $\rm L/I$ & $\rm V/I$\\
& (MHz)&  & & & \\
\hline
J1022+1001 & 1380 & 73.9$\pm$0.2 & --18.0$\pm$0.2  & 25.9$\pm$1.5 , 73.8$\pm$1.5  & --14.6$\pm$1.6 \\ 
J1022+1001 & 4500 & 3.2$\pm$0.3 & 6.9$\pm$0.4 & 26.5$\pm$0.7 & 6.0$\pm$1.3   \\ 
J1713+0747 & 1380 & 36.6$\pm$0.5 & 0.9$\pm$0.5 & 36.3$\pm$0.9 & --0.5$\pm$1.5\\
J1713+0747 & 4500 & 30.1$\pm$0.6 & --1.4$\pm$0.8& 34.5$\pm$0.8 & --0.6$\pm$1.2\\
B1855+09 & 1380 & 6.6$\pm$0.1 & --3.5$\pm$0.2 & 27.3$\pm$0.9 & --4.6$\pm$1.7 \\ 
\hline
\label{tb:pol}
\end{tabular}
\end{footnotesize}\end{center}
\begin{tablenotes}
\small
\item For PSR~J1022+1001 at 1.38 GHz the peak fractional linear polarization for leading and trailing components are listed separately.
\end{tablenotes}
\end{table*}

\begin{figure*}
    \begin{center}
    \hspace{-1.cm}
    \subfigure[]
{
\includegraphics[width=3.7cm,trim={ 0.3cm 0.0cm 0.4cm 0cm},clip]{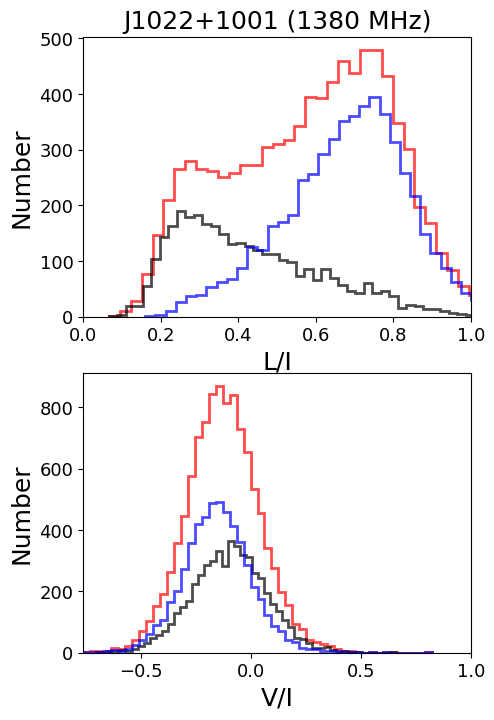} 
\label{fig:pa_J1022_L}
}
\subfigure[]
{
\includegraphics[width=3.45cm,trim={ 0.3cm 0.0cm 0.4cm 0cm},clip]{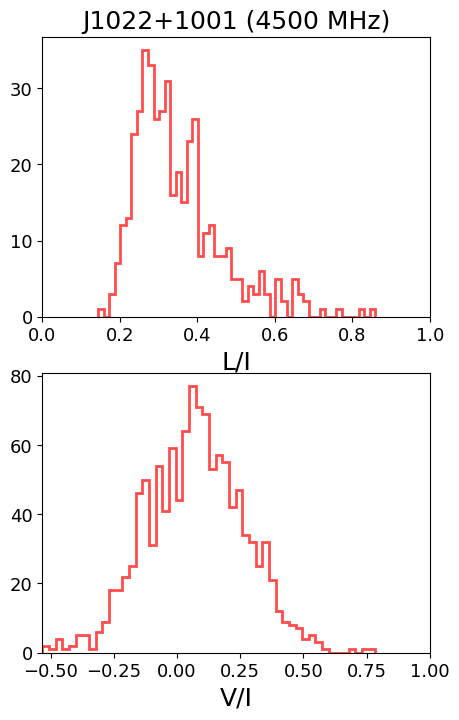} 
}
 \subfigure[]
{
\includegraphics[width=3.6cm,trim={ 0.3cm 0.0cm 0.4cm 0cm},clip]{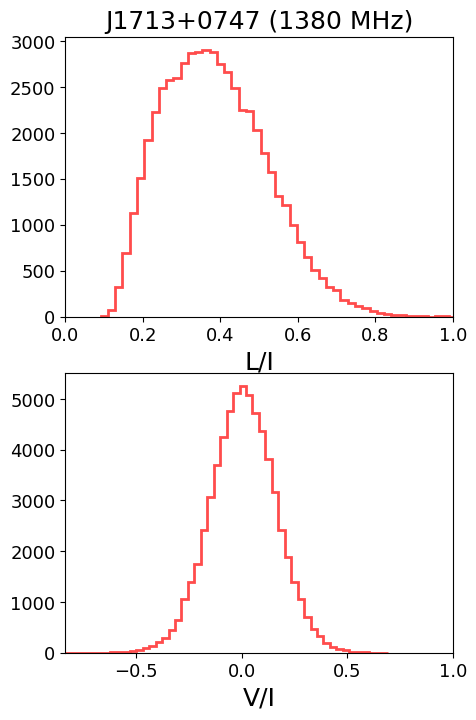} 
}
\subfigure[]
{
\includegraphics[width=3.4cm,trim={ 0.3cm 0.0cm 0.4cm 0cm},clip]{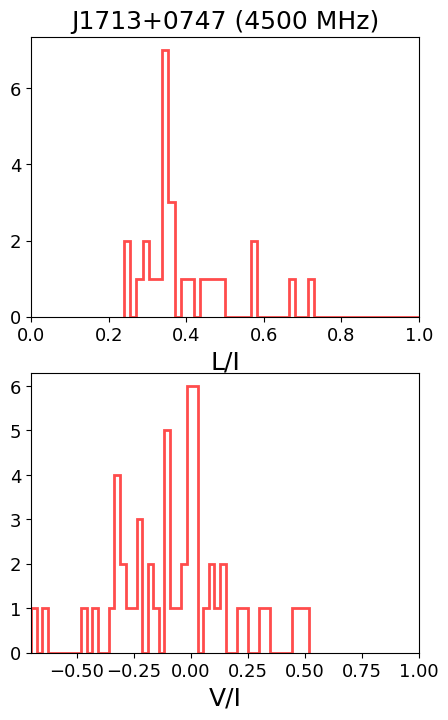} 
}
    \subfigure[]
    {
\includegraphics[width=3.45cm,trim={ 0.3cm 0.0cm 0.38cm 0cm},clip]{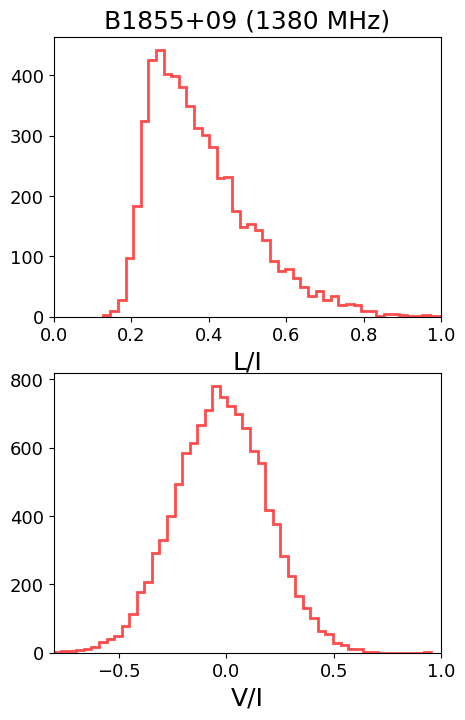} 
}
\caption{
Histograms of linear polarization (top) and circular polarization (bottom) for PSR~J1022+1001 at 1.38 GHz (a) and 4.5 GHz (b), PSR~J1713+0747 at 1.38 GHz (c), and 4.5 GHz (d), and  PSR~B1855+09 at 1.38 GHz (e).
For PSR~J1022+1001 at 1.38 GHz, the histograms of single--pulses associated with the leading (black) and trailing (blue) components are also shown separately. Single pulses associated with the trailing component show a higher fractional linear polarization than those associated with the leading component. 
}
\label{fig:linpol_single}
\end{center}
\end{figure*}

\begin{figure*}
    \begin{center}
    \hspace{-1.cm}
    \subfigure[]
{
\includegraphics[width=3.6cm,trim={ 0.0cm 0.0cm 0.5cm 0cm},clip]{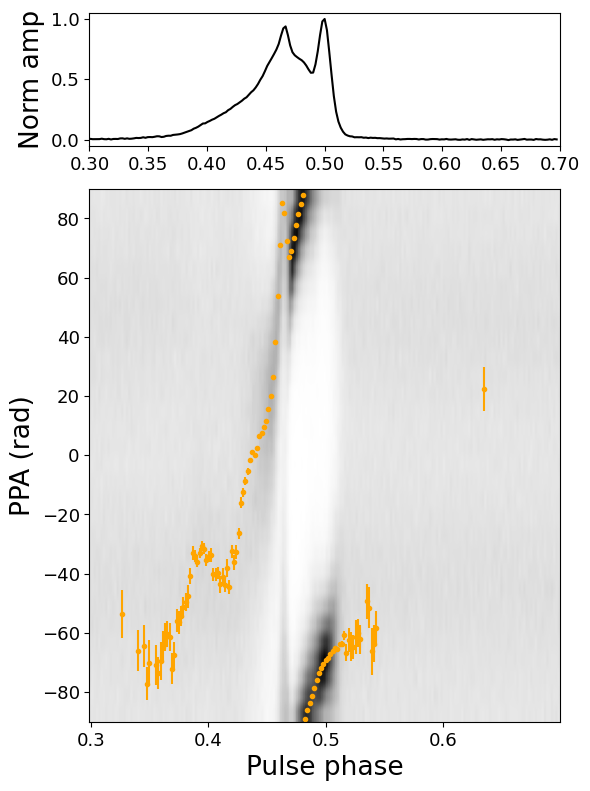} 
}
\subfigure[]
{
\includegraphics[width=3.45cm,trim={ 1.0cm 0.0cm 0.5cm 0cm},clip]{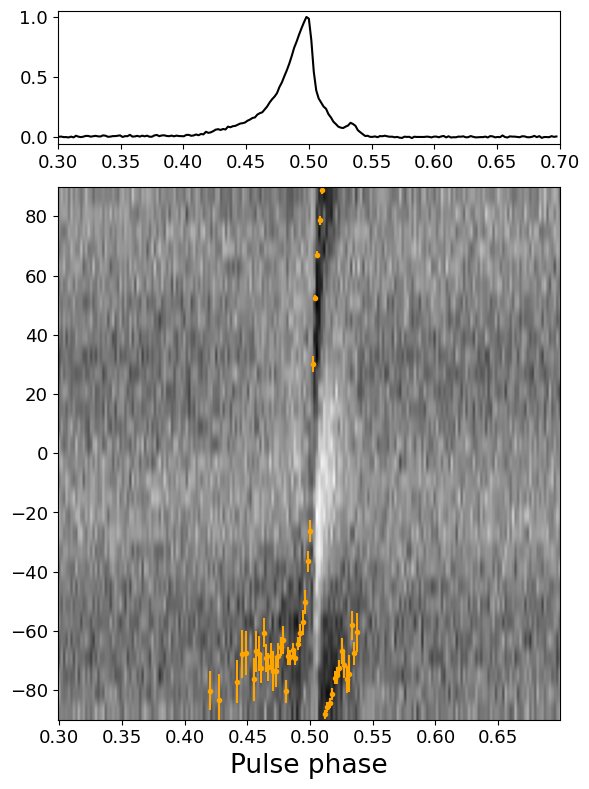} 
}
  \subfigure[]
    {
\includegraphics[width=3.45cm,trim={ 1.0cm 0.0cm 0.5cm 0cm},clip]{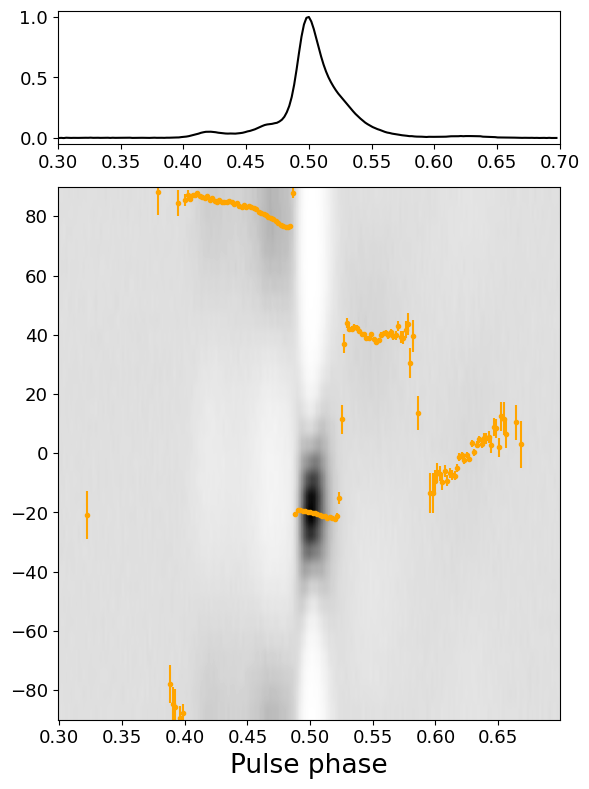} 
}
 \subfigure[]
    {
\includegraphics[width=3.45cm,trim={ 1.0cm 0.0cm 0.5cm 0cm},clip]{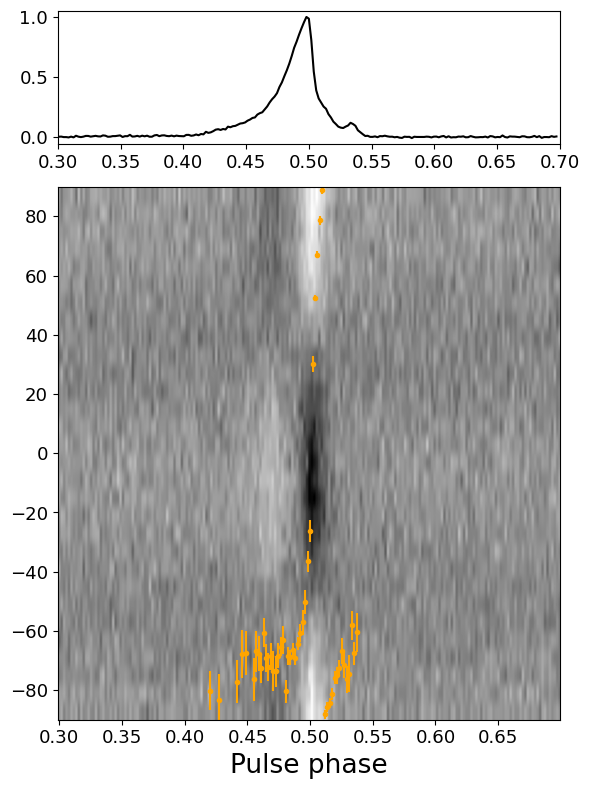} 
\label{fig:pahist_single_J1713_4500}
}
    \subfigure[]
    {
\includegraphics[width=3.5cm,trim={ 0.8cm 0.0cm 0.5cm 0cm},clip]{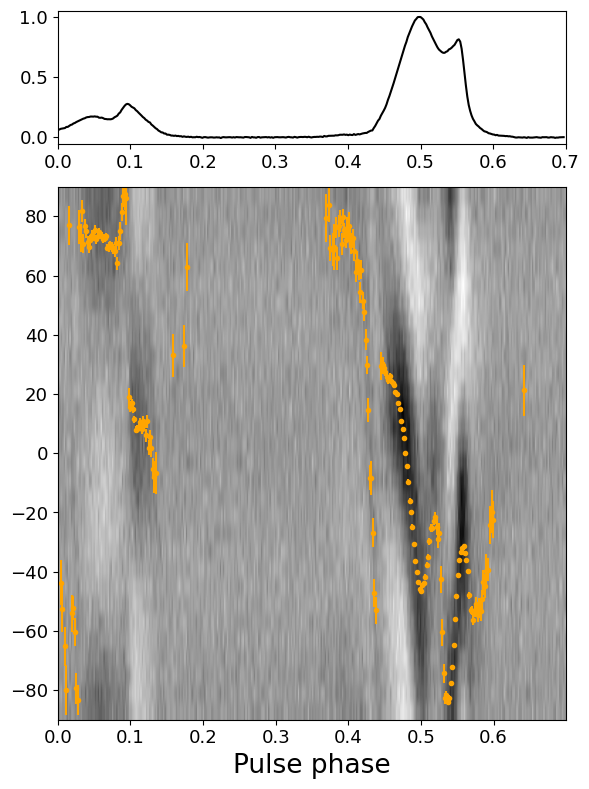} 
}
\caption{
The total intensity average profile of the full dataset (top) and the single-pulse PPA distribution (bottom grey scale) of for PSR~J1022+1001 at 1.38 GHz (a) and 4.5 GHz (b), PSR~J1713+0747 at 1.38 GHz (c), and 4.5 GHz (d), and  PSR~B1855+09 at 1.38 GHz (e).
The PPA of the average profile (orange) is also marked.
}
\label{fig:pahist_single}
\end{center}
\end{figure*}

\subsection{Pulse phase jitter}

\begin{figure*}
    \begin{center}
    \hspace{-1.cm}
    \subfigure[]{
\includegraphics[width=3.6cm,trim={ 0.2cm 0.0cm 0.3cm 0.2cm},clip]{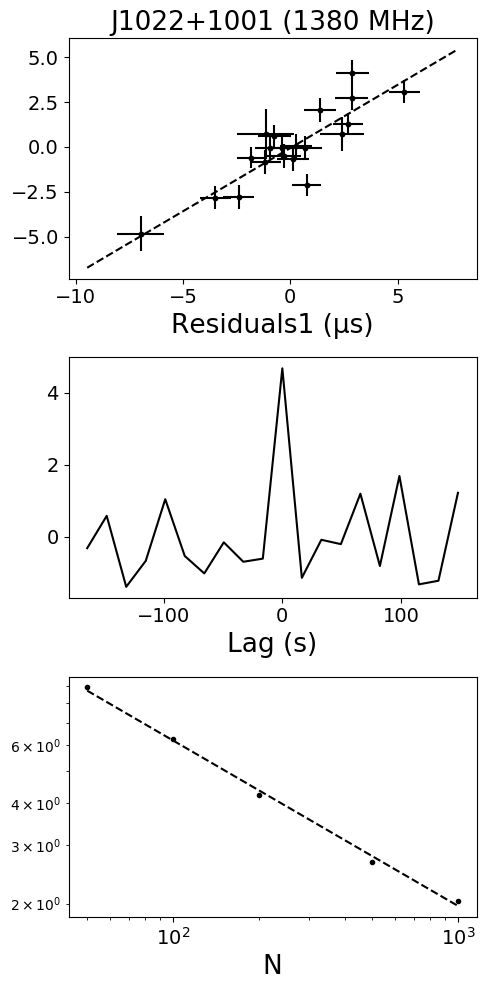} 
}
\subfigure[]{
\includegraphics[width=3.55cm,trim={ 0.2cm 0cm 0.3cm 0.2cm},clip]{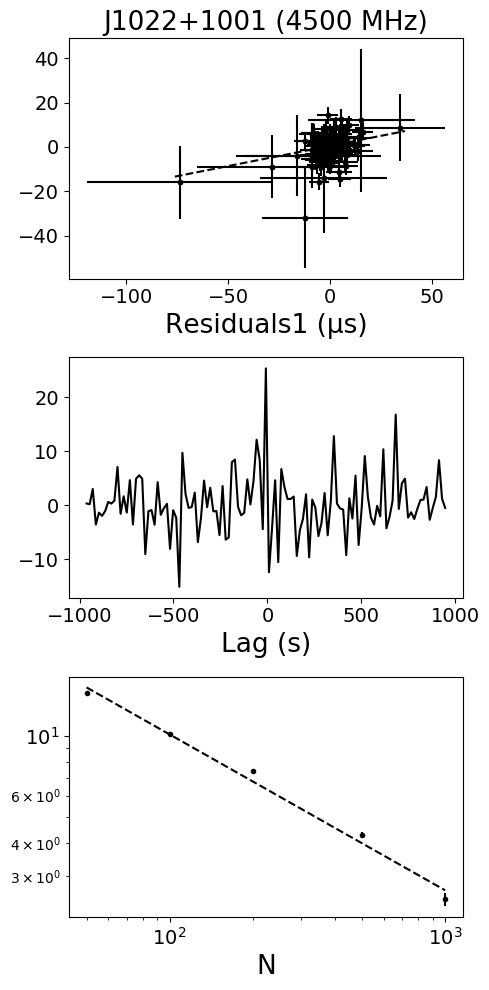}
}
\subfigure[]{
\includegraphics[width=3.55cm,trim={ 0.2cm 0cm 0.3cm 0.2cm},clip]{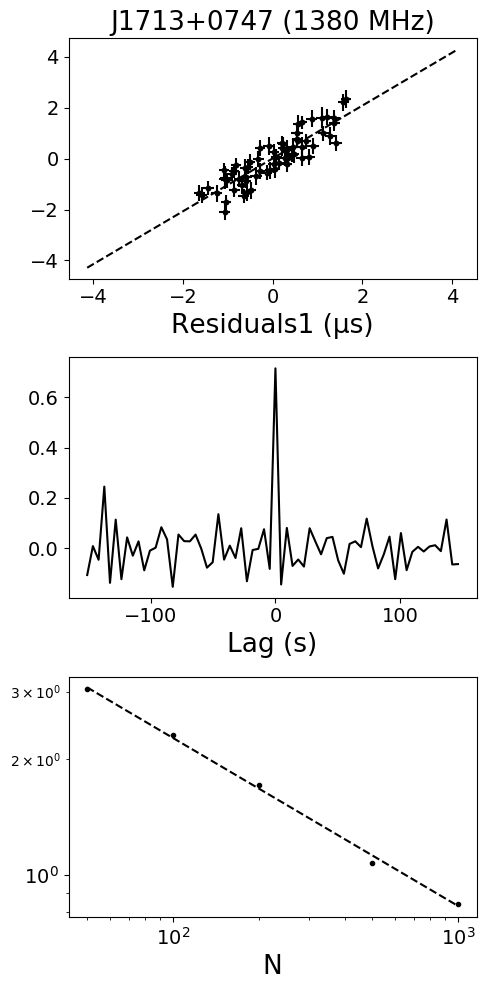} 
}
\subfigure[]{
\includegraphics[width=3.55cm,trim={ 0.2cm 0.1cm 0.3cm 0.2cm},clip]{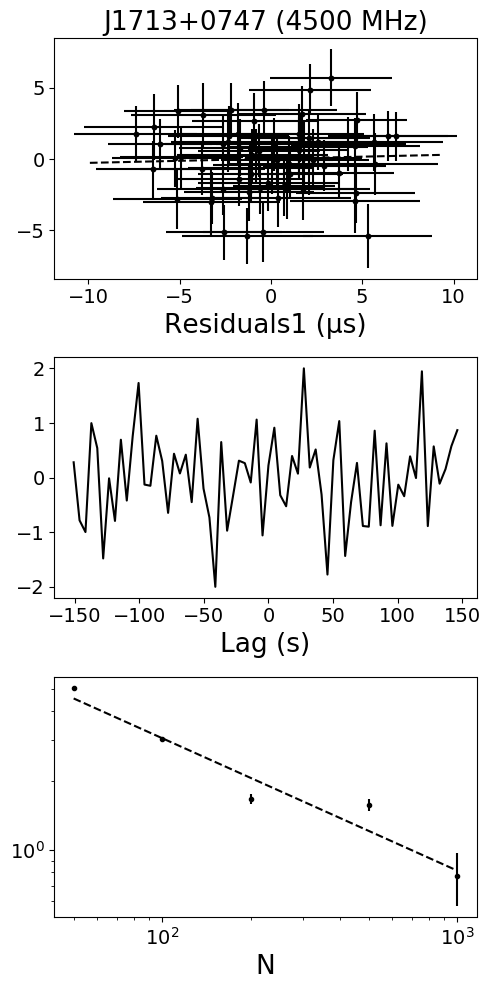}
\label{fig:jitter_J1713_4500}
}
\subfigure[]{
\includegraphics[width=3.55cm,trim={ 0.2cm 0.1cm 0.2cm 0.2cm},clip]{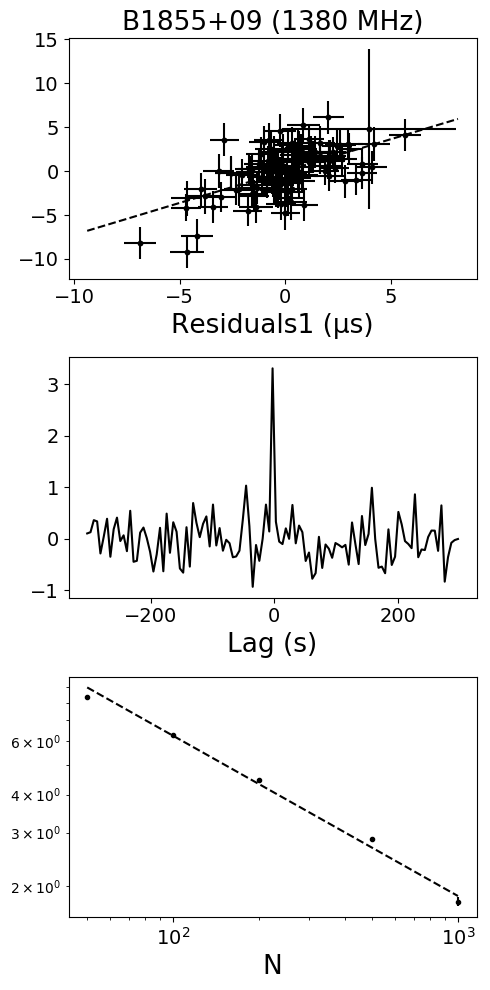}
}
\caption{
Correlation between TOAs of top and bottom halves of the band for 1000-pulse averages along with the best-fit line (top), the cross correlation function (middle), and the rms timing residuals from jitter vs the number of pulses averaged (bottom) for PSR~J1022+1001  at 1.38 GHz (a) and 4.5 GHz (b), PSR~J1713+0747 at 1.38 GHz (c), and 4.5 GHz (d), and  PSR~B1855+09 at 1.38 GHz (e). 
In the bottom plots the best-fit line to jitter noise is shown by the dashed line.
The correlation is negligible for both PSRs J1022+1001 and J1713+0747 at 4.5 GHz for 1000-pulse averages likely due to radiometer noise in profiles. 
}
\label{fig:jitter}
\end{center}
\end{figure*}

\begin{figure}
    \begin{center}
    \hspace{-1.cm}
    \includegraphics[width=7.0cm,trim={ 0cm 0.0cm 0.0cm 0cm},clip]{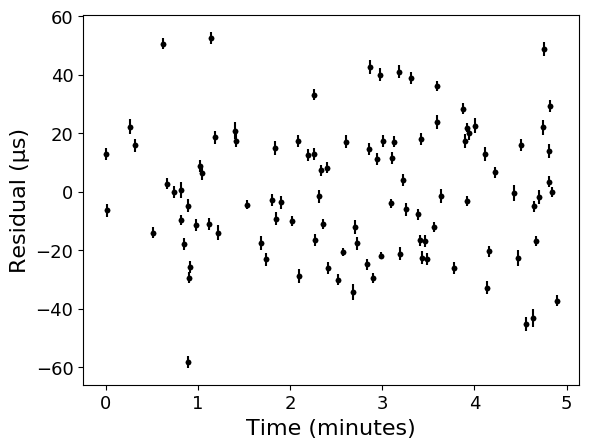}
\caption{
The rms timing residuals of PSR~J1022+1001 single--pulses of $\rm S/N > 25$ at 1.38 GHz. 
}
\label{fig:timingbright}
\end{center}
\end{figure}

As demonstrated by \citet{sc12}, timing residuals due to pulse--to--pulse jitter are expected to be correlated across radio frequency.
For average profiles formed with $N$ pulses,  the root-mean-square (rms) residuals due to jitter noise, $\sigma_J(N)$, is quantified as the quadrature difference between the measured rms timing residuals, $\sigma_{\rm obs}(N)$, and the expected rms timing residuals from radiometer noise only, $\sigma_{\rm rad}(N)$, such that \citep{sc12}
\begin{equation}
\sigma_J(N)^2 = \sigma_{\rm obs}(N)^2 - \sigma_{\rm rad}(N)^2.
\label{eq:jit}
\end{equation}
In order to determine the effect of jitter, composite profiles were formed with $N=$ 50, 100, 200, 500, and 1000 pulses averaged together.
For these various $N$ values, $\sigma_{\rm obs}(N)$ was calculated from the timing residuals of the measured TOAs (obtained as described in Section~\ref{sec:obs}) and $\sigma_{\rm rad}(N)$ was calculated from simulated profiles of similar S/N levels as the composite profiles \citep{dr93}.
The simulated profiles were constructed by adding white noise to a noise-free template with signal-to-noise drawn from the signal-to-noise distribution of the observed composite profiles.
We note that the simulated dataset does not contain uncertainties due to profile shape variations or any other phenomena.
The rms residuals due to jitter noise were determined from Equation~\ref{eq:jit} for various $N$ values. 

Timing residuals due to jitter noise are correlated between sub-bands.
The band may be split into several sub-bands and the correlation may be calculated between adjacent sub-bands \citep{pbs21}.
Following \citet{sc12}, we calculate the rms due to jitter by correlating timing residuals from the top and bottom halves of the band as
$\sigma_{j}=\sqrt{\rm{CCF(0)}}$. 
Here, CCF(0) is the  zero--lag value of the cross--correlation function. 
A different template profile was used for each subband.

\begin{table*}
\begin{center}
\begin{footnotesize}
\caption{Phase jitter analysis: Pulsar name, frequency, the slope of the best fit to the timing residual correlation for 1000--pulse averages ($m$ -see the top panel of Figure~\ref{fig:jitter}), the corresponding zero--lag value of the CCF (CCF(0)), rms residuals due to jitter from the CCF ($\sigma_{\rm j,CCF}$),  measured rms timing residuals ($\sigma_{\rm obs}$), the reduced chi-square of the timing residuals ($\chi_r^2$), rms residuals due to radiometer noise ($\sigma_{\rm rad}$), rms residuals due to jitter ($\sigma_{\rm j}$) from the quadrature difference between $\sigma_{\rm obs}$ and $\sigma_{\rm rad}$, jitter scaling parameters (A, $\beta$ -- see text), and  $\sigma_{\rm j}$   scaled to one hour ($\sigma_{\rm j,1\,hr}$). 
}
\begin{tabular}{lllllllllll}
\hline
Pulsar & Frequency & $m$ & CCF(0)& $\sigma_{\rm j,CCF}$ & $\sigma_{\rm obs}$&  $\chi_r^2$& $\sigma_{\rm rad}$ & $\sigma_{\rm j}$ & (A, $\beta$)& $\sigma_{\rm j,1\,hr}$\\
& (MHz)&  & ($\mu$s$^2$)  & (ns)& (ns)& & (ns)& (ns)& & (ns)\\
\hline
J1022+1001 & 1380 & 0.71 & 4.7 & 2164 &  2086&16.1& 467 & 2033 & (61.10,$-$0.50) & 135$\pm29$\\ 
J1022+1001 & 4500 & 0.18 & 25.3 & 5033 & 4502 & 1.69& 3764 &2469 & (146.4,$-0.58$) & 117$\pm$52\\ 
J1713+0747 & 1380 & 1.04 & 0.72 & 846 & 851 &49.9 &  121 & 842 & (17.03,$-$0.44) & 45$\pm$7  \\
J1713+0747 & 4500 & 0.03 & 0.22& 473 &  1939 &1.30 & 1779 &  772 & (43.15,$-$0.57)  &17$\pm$22\\
B1855+09 & 1380 & 0.80 &3.31 & 1819 &  2003 & 7.87 &
914& 1782 & (70.39,$-$0.53) & 60$\pm$25\\ 
\hline
\label{tb:jitter}
\end{tabular}
\end{footnotesize}\end{center}\end{table*}

\newpage
\section{Results}
\label{sec:res}

In this section we first present overall single--pulse total intensity, polarimetry, and jitter results and then present the results for individual pulsars in subsections.

\textit{Single--pulse properties:}
Single pulses of $\rm S/N>5$ are detected for PSRs J1022+1001 and J1713+0747 at 1.38~GHz and 4.5~GHz, and for B1855+09 at 1.38~GHz.  The 2.3 GHz data set of B1855+09 does not show detectable single pulses. To the best of our knowledge, this is the first report of the detection of single pulses at frequencies $>2$ GHz in MSPs.

Figure~\ref{fig:singleprofs} shows a single-pulse sequence (a stack of individual pulses in time vs pulse--phase), the total intensity and polarization profiles (see Section~\ref{sp_pol} for polarimetry) of averaged data and the three highest S/N individual single pulses from each data set.
Figure~\ref{fig:ampdist} shows the histograms of peak pulse amplitude distribution and the S/N distribution of single pulses of PSRs J1022+1001, J1713$+$0747, and B1855+09 at frequencies 1.38~GHz and 4.5~GHz.

Figure~\ref{fig:width_vs_snr} shows the single--pulse equivalent width vs S/N. 
Overall, single pulses with  higher S/N seem to be narrower than the ones with smaller S/N at 1.38 GHz, as shown in Figure~\ref{fig:width_vs_snr}. 
In general, Gaussians or a series of Gaussians are fit to pulse profiles to estimate the width. 
The equivalent width method is better for automated calculation for many pulses, where goodness of fit is difficult to ascertain for each pulse, and to facilitate width measurements for weak pulses.
At low S/N, the true width may be difficult to measure resulting in large width estimates. 
In Figure~\ref{fig:width_vs_snr}, we plot width measurements determined with > $3\sigma$ significance.

\textit{Polarimetry:}
Figure~\ref{fig:linpol_single} shows the histograms of single--pulse linear and circular polarization fraction at the peak. 
The bias in L has been removed according to \citet{ev01}.
Table~\ref{tb:pol} lists the linear and circular polarization fraction of the average profiles and the peak of the distributions of single-pulse polarization fractions shown in Figure~\ref{fig:linpol_single}.
The error on the polarization fraction of the average profile are obtained by using the standard deviation of I, L, and V across the profile and applying error propagation. 

Phase resolved histograms of the PPA for single--pulse data were obtained using \texttt{psrspa} routine in \textsc{psrchive}.
Figure~\ref{fig:pahist_single} shows the logarithm distribution of single-pulse PPA as a function of pulse phase along with the PPA of the average profile. The histogram is weighted by the total intensity.

\textit{Pulse phase jitter:}
Figure~\ref{fig:jitter} shows the correlation between the timing residuals from the top and bottom halves of the band, for 1000-pulse averages,  along with the best-fit line whose slope should be ideally be unity.
The slopes of the best-fit lines for the three pulsars are listed in Table~\ref{tb:jitter}.
Table~\ref{tb:jitter} also lists the zero lag value of the cross--correlation function CCF(0), rms timing residuals $\sigma_{\rm obs}$ for 1000-pulse averages, the reduced chi-square of the timing residuals ($\chi_r^2$), the TOA error from radiometer noise $\sigma_{\rm rad}$,  rms residuals due to jitter calculated from Equation~\ref{eq:jit},  the best--fit parameters of the line $\sigma_J(N)$ vs $N$, and rms residuals due to jitter scaled to one hour.
The rms residuals due to jitter for one hour averages were determined from the  parameters of the best--fit line to $\sigma_J(N)$ vs $N$ such that $\sigma_J(N)=A\,N^{\beta}$, as shown in the bottom panel of Figure~\ref{fig:jitter}.
The error on the jitter level scaled to one hour is obtained by applying error propagation to $\sigma_J(N)=A\,N^{\beta}$. 
The quoted error bars correspond to $1\sigma$ level.

In the quadrature difference method, measuring the radiometer noise component accurately is important to get a reliable estimate for jitter noise since the datasets at higher frequencies are in general of low S/N.
Higher S/N for these pulsars would not be achievable with any other existing telescope at 4.5 GHz. 
A comparison of the jitter level from the cross-correlation method and quadrature difference ($\sigma_{\rm j,CCF}$ and $\sigma_j$ in Table~\ref{tb:jitter}) between the observed and simulated rms timing residuals  shows that the values are comparable (within a factor of $\lesssim1.1$) in all datasets at 1.38 GHz. The inconsistencies at 4.5 GHz in both datasets of J1022+1001 and J1713+0747 could perhaps be because the radiometer noise component is underestimated or because jitter is decorrelated across the band \citep[e.g.][]{pbs21}. 

\subsection{PSR J1022+1001}

As shown in Figure~\ref{fig:singleprofs}, PSR~J1022+1001 shows a double--peaked profile with a highly linearly polarized ($\approx74\%$) trailing component at 1.38 GHz, as also noted previously \citep{hbo04,Liu2015}. At 4.5 GHz the  leading component has reduced in amplitude with a more prominent trailing component which is only $\sim 3\%$ linearly polarized. 
While the percentage of circular polarization is $~18\%$ at 1.38 GHz (under the leading component), it is almost negligible at 4.5 GHz.
The PPA of the average profile follows an S-shaped swing with a "notch" at the first peak (Figure~\ref{fig:pa_J1022_L}) as also reported by \citet{kxc99}.
This pulsar shows  long--term profile variations \citep{kxc99}, which have been attributed to improper polarization calibration in the past \citep{v13}, even though more recent studies suggest that intrinsic effects may likely be at play \citep{Liu2015,pbc21}.
By using the measured flux densities given in Table~\ref{tb:obs}, we measure a spectral index of $-1.43\pm0.24$, consistent (within errors) with the published value of $-1.7\pm0.1$ \citep{kxl98}. 
Since the observed flux density of this pulsar is known to vary significantly due to diffractive scintillation \citep{Liu2015,Shannon2014}, flux density measurements over multiple epochs are needed for accurate spectral index measurements.

We do not see evidence for giant pulses. However, the data set consists of very bright pulses of (S/N$\approx30$), which are $\approx\times5$ the mean S/N. 
At 4.5 GHz, some bright single pulses appear at phase $\approx 0.53$ (phase $\approx0.03$ away from the main component) as shown in Figure~\ref{fig:singleprofs}.

The brightest single pulses of PSR~J1022+1001 at 1.38 GHz are highly linearly polarized, as also noted in other studies \citep{Liu2015}.
We find that the single--pulse linear polarization distribution of PSR~J1022+1001 at 1.38 GHz shows two peaks (see Figure~\ref{fig:linpol_single}).
Single pulses associated with the leading component of the average profile tend to have a low linear polarization fraction (average $L/I\approx0.3$) compared to the single pulses associated with the second peak (average $L/I\approx0.7$).
As evident from Table~\ref{tb:pol} and Figure~\ref{fig:linpol_single}, in general, the fraction of linear polarization is reduced at higher frequencies. 
The brightest single pulses at 4.5 GHz (of $\rm S/N>5$ which we consider in our analysis) show higher fractional $L/I$ compared to the average profile.

The rms timing residual due to pulse--to--pulse jitter at 1.38 GHz, scaled to one hour, is $135\pm29$ ns. This is consistent (within the 2-$\sigma$ error) with the $67\pm9$ reported by \citet{Feng2020} or $\approx 700$ ns for 1-min integration ($\approx 90$ ns for one hour) reported by \citet{Liu2015} at 1.38 GHz, but lower than the \citet{Shannon2014} value of $290\pm15$ ns and $269\pm4$ ns of \citet{Lam2019}.
The level of jitter at 4.5 GHz, scaled to one hour, is 117$\pm$52 ns, which is consistent with that at 1.38 GHz within errors. 

Following \citet{Liu2015} and \citet{Feng2020}, we measured the rms timing residuals of single pulses from the leading and trailing component separately.
Considering all single pulses with $\rm S/N>5$ results in rms timing residuals of $\approx70\, \mu$s and $\approx43 \, \mu$s for leading and trailing components respectively.
For TOA calculation, the template profile was created by averaging over all pulses of $\rm S/N>5$ within each component.
\citet{Liu2015} finds rms timing residuals of $4.3\,\mu$s for averages of 30 trailing component sub-pulses of $\rm S/N>5$, which converts to $4.3\times\sqrt{30}\approx23\,\mu$s at the single--pulse level. 
The reason for higher rms timing residuals in our dataset could be the higher levels of jitter compared to \citet{Liu2015} possibly due to the pulsar showing variable jitter noise or the difference in radiometer noise.

We also measured the rms timing residuals of single pulses within S/N ranges of $5<\rm S/N < 15$, $15<\rm S/N < 25$ and S/N $> 25$.  As shown in Figure~\ref{fig:timingbright}, the rms timing residual using the brightest pulses ($\rm S/N>25$) is $\approx 22 \, \mu$s. 
For TOA calculation, the template profile was created by averaging over all pulses within each S/N range.
Scaling the rms of $\approx2.09\,\mu$s from 1000-pulse averages (Table~\ref{tb:jitter}) simply as $\sigma_{obs} \propto 1/\sqrt{N_p}$ \citep[e.g.][]{sc12}, where $N_p$ is the number of pulses in the average profile, the expected single--pulse RMS is $\approx 66\,\mu$s. So timing just the brightest pulses gives a factor of $\approx 3$ improvement. 
The vast majority of pulses (99\%) within $\rm S/N>25$ range appear at the trailing component.
Using single pulses within the S/N ranges of $5<\rm S/N < 15$ and  $15<\rm S/N < 25$ yields rms timing residuals of 69 and 74 $\mu$s respectively.
The higher rms residuals in pulses within the $15<\rm S/N<25$ range could be due to a small percentage (3.7\%) of pulses occurring at the leading component while the rest occur at the trailing component.  
When these trailing edge pulses are removed, the rms timing residuals within $15<\rm S/N<25$ reduce to 28 $\mu$s.

\subsection{PSR J1713+0747}

PSR J1713+0747 is a binary pulsar which shows relativistic effects \citep{camilo94,splaver04}.
PSR~J1713+0747 shows a multi--component average profile with a significant linear polarization ($\approx 37\%$) at the peak at 1.38 GHz, consistent with previous studies \citep[e.g.][]{Yan11}. 
This is reduced but still significant ($\approx 30\%$) at 4.5 GHz.
We measure a spectral index of $-1.69\pm0.38$ using the flux densities obtained at 1.38 GHz and 4.5 GHz (see Table~\ref{tb:obs}), which is consistent (within errors) with the published value of $-1.5\pm0.1$ \citep{kxl98}.
The more prominent outer components at 4.5 GHz compared to at 1.38 GHz (at phases $\approx$0.42 and 0.47), also reported in \citet{Kijak1997}, are evident in our data (see Figure~\ref{fig:singleprofs}).
The faint emission component at phase $\approx$0.63 at 1.38 GHz is also more prominent at 4.5 GHz.
Previous studies have reported a jitter contribution of $\approx35\pm0.8$ ns to rms timing residuals residuals in one hour of integration of this pulsar \citep{Shannon2014,Liu2016}, consistent (within the $2\sigma$ error) with the $45\pm7$ ns that we find at 1.38 GHz.
At 4.5 GHz, the lack of correlation (see the top two panels of Figure \ref{fig:jitter_J1713_4500}) and the large error in the quadrature difference method point to jitter being not constrained.

Although, the data set consists of very bright pulses of S/N$\approx20$, we do not see evidence for giant pulses.
The fraction of single pulses detected at 4.5 GHz is very small, which stops us from performing detailed analysis.

The PPA of the average profile shows 90$^\circ$ jumps (see Figure~\ref{fig:pahist_single}) that indicate the presence of OPMs \citep{Xilouris98}.
Indeed OPMs have been detected in this pulsar at 1.38 GHz in Large European Array for Pulsars (LEAP) data \citep{Liu2016}.
A faint offset in the PPA distribution of single pulses at phase $\approx0.5$ is noted in Figure~\ref{fig:pahist_single}, hinting at an OPM for this pulsar. 

\subsection{PSR B1855+09}

PSR~B1855+09 is a relativistic binary which has both main pulse (MP) and an interpulse (IP), with both showing a double--peaked structure and weak ($\approx7\%$ and $\approx2\%$ in the MP and IP respectively) linear polarization consistent with previous studies \citep{kxl98}. 
The PPA of the average profile in the MP does not follow an S--shaped swing. The discontinuities in the PPA curve of the IP is non--orthogonal and consistent with \citep{Xilouris98} even though some studies have noted orthogonal transitions \citep{Yan11}. 
Based on the measured flux densities at 430 MHz and 1.38 GHz, we measure a spectral index of 
$-0.6\pm0.1$, slightly flatter than the published value of $-1.3\pm0.2$ \citep{kxl98}. 
The flux density measurement at 2 GHz is unusually low, likely due to the very low S/N of the data set or scintillation, and therefore has not been included in the spectral index measurement.
We also note that our data are only polarization calibrated, and therefore these flux density measurements may not be precise. 
Therefore we again emphasise that multi--epoch flux density measurements may give a better estimate of the the spectral index.
Also note that the MSP spectral index is not well constrained \citep{Bates2013, Kuniyoshi2015}.
We find that the rms timing residual due to jitter, scaled to one hour, is $60\pm25$ ns ($70\pm11\,\mu$s at the single--pulse level). 
This is roughly consistent within the $3\sigma$ 
confidence interval of $90 - 115 \,\mu$s published in \citet{Lam2019}. 

Single pulses of PSR~B1855+09 at 1.38 GHz also show  higher fractional $L/I$ compared to the average profile.
We also note that even though the flux density at 430 MHz is high, the expected S/N of a single pulse is low and therefore single pulses are not detected at these low frequencies possibly due to the high sky temperature. 
For Arecibo observations (parameters listed in Table~\ref{tb:obs}), the expected S/N, for B1855+09 at 430 MHz, calculated from the radiometer equation, 
\begin{equation}
 S/N=\frac{G\sqrt{N_{p}W\,\Delta\nu}\,S_{\rm peak}}{T_{\rm sys}}  
\end{equation}
is $<5$. Here $S_{\rm peak}$ is the peak flux density (estimated using the mean flux density and duty cycle) and we assume that the single-pulse width is $W\approx0.17$ ms, which is the mean single--pulse equivalent width of B1855+09 at 1.38 GHz. 

A few single pulses associated with the IP are detected with low S/N (example shown in the bottom panel of Figure~\ref{fig:singleprofs}). However, due to their low S/N, only single--pulses associated with the MP are used for detailed analysis.

\section{Conclusion and discussion}
\label{sec:sum}

We have studied the total intensity properties, single--pulse polarimetry and pulse phase jitter of MSPs J1022+1001, J1713+0747, B1855+09 at multiple frequencies. On the single--pulse front, we do not detect giant pulses, but we still find bright single pulses at 1.38 GHz for all three MSPs. 
Overall, single pulses with higher S/N seems to be narrower than the ones with smaller S/N at 1.38 GHz, as shown in Figure~\ref{fig:width_vs_snr}. 

As evident from Table~\ref{tb:pol}, in general, the fraction of linear polarization in the average profile is reduced at higher frequencies.
This may be consistent with similar findings of normal pulsars that show that average profiles are depolarized at higher frequencies \citep[e.g.][]{Xilouris1996,jkw06}.
This is because high frequency emission arises closer to the pulsar surface where the magnetic field deviates from a dipole field resulting in superposition of modes and hence significant depolarization \citep[e.g.][]{M97,wwh15,vwt17}. 
\citet{dhm15} reported this trend for PSRs J1022+1001, J1713+0747 and B1855+09 using 1.38 GHz and 3 GHz observations.
Our results confirm that this trend continues at 4.5~GHz for PSRs J1022+1001 and J1713+0747.
Single pulses from PSR~J1022+1001 at 4.5 GHz and PSR B1855+09 at 1.38 GHz show high fractional linear polarization than the average profile. This could be because bright pulses show more fractional polarization or due to depolarization from averaging.

In our data set the fraction of single pulses detected at 4.5~GHz is very small, which prevents us from performing detailed analysis. 
Sensitive observations are needed to better understand single--pulse phenomena at higher frequencies. For example, the expected S/N at 4.5 GHz for a FAST--like telescope (even though FAST cannot observe at frequencies > 2 GHz with current system), with a gain of $ \rm G=18 \rm \, K/Jy$, $\rm T_{sys}=20$ K, $\rm BW=800$ MHz \citep{li16}, would be a factor of $\approx$ 6 better than our data.
Upcoming facilities with high sensitivity and frequency coverage should facilitate multi--frequency single--pulse studies of pulsars.


The jitter levels ($\sigma_j$) we find are consistent for PSRs J1022+1002 \citep{Liu2015,Feng2020,pbs21}, PSR~J1713+0747 \citep{Shannon2014,Liu2016} and PSR B1855+09 \citep{Lam2019} at 1.38 GHz.
But the fact that PSR J1022+1001 has lower values of jitter recorded in other studies \citep{Shannon2014, Lam2019} could indicate variable jitter noise.
The contribution to the pulsar noise budget from jitter at 4.5 GHz and 1.38 GHz are consistent (within the $2\sigma$ error) for PSR~J1022+1001.
The level of jitter from the cross-correlation method ($\sigma_{\rm j,CCF}$) and the quadrature difference method ($\sigma_j$ in Table~\ref{tb:jitter}) are comparable at 1.38 GHz.
At 4.5 GHz either the radiometer noise component is underestimated or jitter is decorrelated across the band \citep[e.g.][]{pbs21}.
Jitter noise is expected to be lower at higher frequencies due to narrower profiles and lower flux densities, though previous studies have shown that not all pulsars follow this trend \citep{Shannon2014, Lam2019}.
While some MSPs used in PTAs show levels of jitter as low as tens of ns per hour, others (which have large pulse widths and small periods) show levels of jitter amounting to hundreds of ns per hour \citep{Shannon2014, pbs21}.
\citet{Lam2019} has found that 43 pulsars out of a sample of 48 shows significant jitter and that 30 of those show significant frequency dependence. 
The general notion is that jitter noise should be detectable in all MSPs if observed with adequate sensitivity \citep{Shannon2014}, and therefore accurately including jitter in noise modeling is important in high-precision pulsar timing.
It also represents a fundamental lower limit on the timing precision achievable for individual MSPs.

We also find that timing only the brightest pulses of PSR J1022+1001 results in an improvement in rms timing residuals by a factor of $\approx$3 than what is expected by scaling timing properties derived from all single pulses.
However, given relative paucity of such bright pulses, the timing may not be improved when compared to fully integrating all available data in a fixed observation time.
We note that previous studies have found that selective timing does not yield a significant improvement in timing \citep{osb14,Feng2020}. The first of these studies reported reduced $\chi^2$ of the timing residuals when selectively integrating only weaker pulses, as well as a minor improvement to timing when using brighter pulses. The reason for our improvement in timing could be because we only include the brightest pulses with a large trailing component and remove the pulses that have a large leading component and thereby selecting only the brightest pulses which appear within the narrow phase range of the trailing component.  
Our very short data set prevents us from timing using integrated profiles of selected bright pulses, but this is an important avenue for further exploration. Therefore, further investigation is needed to see if selective timing yields an improvement for integrated profiles formed with bright pulses within a given component of the profile.

\section*{Data availability}

The data underlying this article will be shared on reasonable request to the corresponding author.

\section*{acknowledgement}
The authors thank Hector Hernandez, Phil Perrillat and other Arecibo Observatory staff for help with scheduling, observation support and data quality checks.
The Arecibo Observatory is a facility of the National Science Foundation operated under cooperative agreement by the University of Central Florida in
alliance with Yang Enterprises, Inc. and Universidad Metropolitana.
The authors also thank the anonymous referee for the constructive remarks on this manuscrpt.
NP, BPP, and MAM are members of the NANOGrav Physics Frontiers Center, supported by NSF award \#1430284 and 
\#2020265.

\end{document}